\documentclass{PoS}

\title{Meson masses at large $\mathbf{N_c}$}

\ShortTitle{Meson masses at large $N_c$}

\author{Gunnar Bali\\
        Institut f\"ur Theoretische Physik, Universit\"at Regensburg,\\
        93040 Regensburg, Germany\\
        E-mail: \email{gunnar.bali@physik.uni-regensburg.de}}

\author{\speaker{Francis Bursa}\\
        Institut f\"ur Theoretische Physik, Universit\"at Regensburg,\\
        93040 Regensburg, Germany\\
        E-mail: \email{francis.bursa@physik.uni-regensburg.de}}

\abstract{We analyze the meson spectrum in $SU(N_c)$ lattice gauge theory.
We use Wilson quarks to measure the pseudoscalar and vector masses
in $SU(2)$, $SU(3)$, $SU(4)$ and $SU(6)$, and extrapolate
to the large--$N_c$ limit. We find that finite--$N_c$ corrections
are small at all values of the quark mass.}

\FullConference{The XXV International Symposium on Lattice Field Theory\\
		 July 30-4 August 2007\\
		 Regensburg, Germany}

\begin{document}

\section{Introduction}
The number of colours $N_c$ is an implicit parameter of QCD. In the large $N_c$ limit the theory becomes simpler~\cite{tHooft}; amongst other features, mesons and glueballs become exactly stable.
The finite--$N_c$ corrections in the pure gauge theory are $\mathcal{O}(1/N_c^2)$. This is also the case in quenched lattice field theory, since in that case fermion loops do not contribute. For full QCD the corrections are $\mathcal{O}(1/N_c)$. Clearly it would be very interesting to evaluate these corrections and establish how large they are in the case of QCD ($N_c=3$).

The corrections can be estimated by carrying out quenched calculations at finite $N_c$. The leading $N_c$ dependence will then give the $\mathcal{O}(1/N_c^2)$ quenched corrections, and an $N_c \to \infty$ extrapolation will give the large $N_c$ limit. Finally this can be compared to unquenched $SU(3)$ results (or indeed directly to experiment) to estimate the $\mathcal{O}(1/N_c)$ unquenched corrections.

Calculation of quantities in the large--$N_c$ limit is also important for attempts to extend the AdS/CFT 
correspondence~\cite{Maldacena} to the case of QCD. This approach, known as AdS/QCD, attempts to learn 
about QCD by studying a five--dimensional theory that is dual to the large--$N_c$ limit of 
QCD~\cite{Witten,PolStr,Csaki,Bab,Krucz,Karch}. The form of this five--dimensional dual (if it exists) is 
unknown, and additional 
information on the QCD side would help to constrain it. Since the duality occurs in the large--$N_c$ limit such information must also be calculated in that limit.

The masses of low--lying glueballs have been calculated on the lattice for $N_c$ up to 
8~\cite{LucTepWen}. The 
finite--$N_c$ corrections were found to be small, reaching only about 10\% in the case of $SU(2)$. The pion mass has been calculated for $N_c=17$ to $23$, with no $N_c$ dependence observed~\cite{NandN}. Other meson masses have not to our knowledege been calculated at large $N_c$.

In this work we present calculations of the pion and rho masses up to $SU(6)$. We describe our methods in the next Section and our results in Section~\ref{results}. We sum up in Section~\ref{conclusions}.

\section{Methods}
We use the lattice software package \emph{Chroma}~\cite{Chroma}, which we adapt to work for arbitrary $N_c$. We use unimproved Wilson fermions, together with the Wilson plaquette action for the gauge fields.

To set the scale, we use the string tension calculations by Lucini \emph{et al.}~\cite{LucTepWen2}. We choose the coupling $\beta=2N_c/g^2=2N_c^2/\lambda$, where $\lambda$ is the 't~Hooft coupling, such that the string tension in lattice units, $a\sqrt\sigma$, is the same for each $N_c$. We use the value $a\sqrt\sigma=0.2093$: for $SU(3)$ this corresponds to $\beta=6.0175$. The values for other $N_c$ are shown in Table~\ref{beta}. Adopting a value of 420 MeV for the string tension, the lattice spacing is then 0.099 fm in each case. The values of $a\sqrt\sigma$ used in the fits in~\cite{LucTepWen2} are very accurate~\cite{LucTepWen}, so the errors on our estimate of the lattice spacing are less than 1\%.

\begin{table}[h]
\begin{center}
\begin{tabular}{|c|c|} \hline
$N_c$ & $\beta$ \\ \hline
2 & 2.4645 \\
3 & 6.0175 \\
4 & 11.028 \\
6 & 25.452 \\ \hline
\end{tabular}
\caption{Value of the coupling $\beta$ for each $N_c$.}
\label{beta}
\end{center}
\end{table}

The volume of our lattices is $16^3\times32$ in lattice units, corresponding to a spatial volume of 
$1.58^3$ fm. Finite--volume effects are expected to decrease with $N_c$, and to be zero at infinite $N_c$ as long as the box is larger than a critical length $l_c$~\cite{KisNarNeu}. This means that we should obtain the correct large--$N_c$ limits for the masses we calculate, despite any finite--volume effects at small $N_c$.

We find that, as expected, the cost of updating a gauge configuration is approximately proportional to 
$N_c^3$, and the cost of inverting a propagator is approximately proportional to $N_c^2$ (the number of conjugate gradient steps required for the inversion is approximately independent of $N_c$). 
For the 
relatively small values of $N_c$ we use, the latter in fact dominates. Furthermore, we find that the correlators become less noisy at larger $N_c$ (see below), so fewer are needed to calculate a mass to a given precision. This reduces the total cost to $\sim N_c$. A heuristic argument supporting this observed reduction in noise is the increase of the degrees of freedom of the statistical system $\propto N_c^2$ at fixed volume.

We calculate local--local and local--smeared correlators on our configurations.
We smear using 0 to 100 iterations of Gaussian smearing, 
$\psi^\prime(x)=\psi(x)+\kappa\sum_{\mu}(U_\mu(x)\psi(x+\mu)+U^\dagger_\mu(x-\mu)\psi(x-\mu))$, with smearing parameter $\kappa=4$. We smear the gauge links using 10 iterations of APE smearing, $U_\mu^\prime(x)=\alpha U_\mu(x)+\sum_{\nu}(U_\nu(x)U_\mu(x+\nu)U_\nu^\dagger(x+\mu)+U_\nu^\dagger(x-\nu)U_\mu(x-\nu)U_\nu(x+\mu-\nu))$, with smearing parameter $\alpha=2.5$

\section{Results}
\label{results}
\begin{figure}[h]
\begin{center}
\input{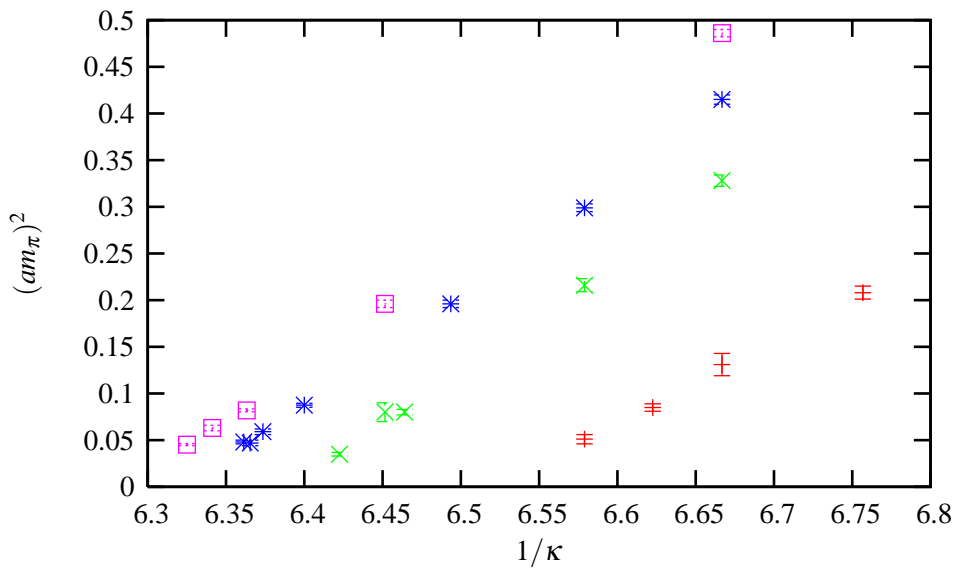}
\end{center}
\vspace{-5mm}
\caption{$(am_\pi)^2$ as a function of $1/\kappa$ for $SU(2)$~($+$), $SU(3)$~($\times$), $SU(4)$~($\ast$) and $SU(6)$~($\Box$).}
\label{kappa_fig}
\end{figure}

We search for the critical value of the hopping parameter, $\kappa_c$, by finding the value of $\kappa$ for which the pion mass $m_\pi$ vanishes. We show a plot $(am_\pi)^2$ as a function of $1/\kappa$ in Fig.~\ref{kappa_fig}. For suffciently small masses the dependence is linear, as expected, and we can extrapolate to zero pion mass. The values of $\kappa_c$ we obtain are shown in Table~\ref{kappa_table}. It is clear that they are rapidly converging to an $N_c=\infty$ value.

\begin{table}[h]
\begin{center}
\begin{tabular}{|c|c|} \hline
$N_c$ & $\kappa_c$ \\ \hline
2 & 0.15327(16) \\
3 & 0.156397(45) \\
4 & 0.158168(39) \\
6 & 0.159060(44) \\ \hline
\end{tabular}
\caption{Critical hopping parameter $\kappa_c$ for each $N_c$.}
\label{kappa_table}
\end{center}
\end{table}

\begin{figure}[h]
\begin{center}
\input{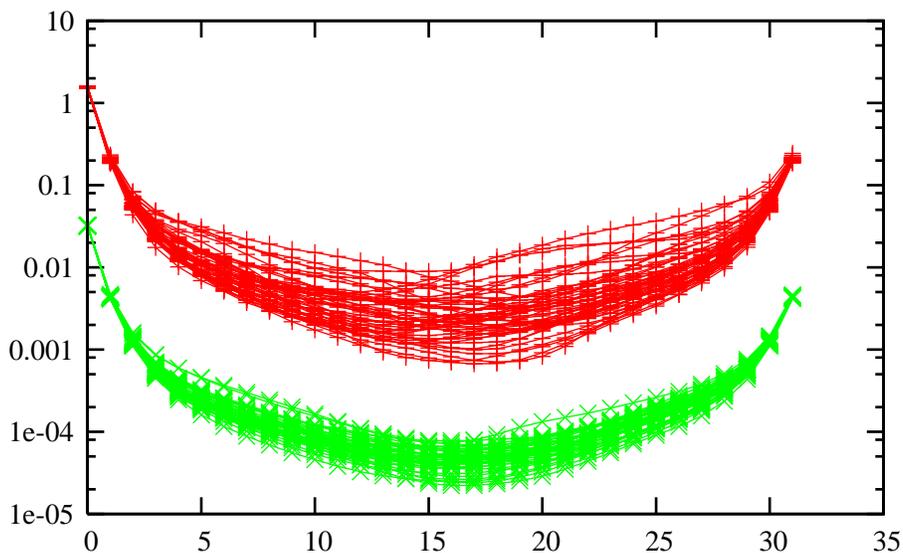}
\end{center}
\vspace{-5mm}
\caption{Point--point pseudoscalar correlators on individual gauge configurations in $SU(3)$ at $\kappa=0.1547$~($+$) and in $SU(6)$ at $\kappa=0.15715$~($\times$). $SU(6)$ results have been shifted vertically for clarity.}
\label{noise_fig}
\end{figure}

As mentioned above, we find that correlators become less noisy as $N_c$ increases. We illustrate this in Fig.~\ref{noise_fig}, where we compare pseudoscalar correlators on individual gauge configurations in $SU(3)$ and $SU(6)$. The pion masses are almost identical, $am_\pi=0.283(5)$ and $0.286(3)$ respectively, but the scatter between individual configurations, a measure of the noise, is about twice as large in $SU(3)$, in agreement with the naive degrees of freedom argument
above.

Since fluctuations decrease as $N_c$ increases, and we use correlators of fluctations to extract masses, 
one might worry that we will be unable to calculate masses in this way in the large--$N_c$ limit. Fortunately this is not the case. The correlators do indeed decrease as $1/N_c^2$, but the fluctations in the correlators decrease at the same rate, so the signal--to--noise ratio remains constant. For a more detailed discussion of this issue see~\cite{LucTepWen}.

\begin{figure}[h]
\begin{center}
\input{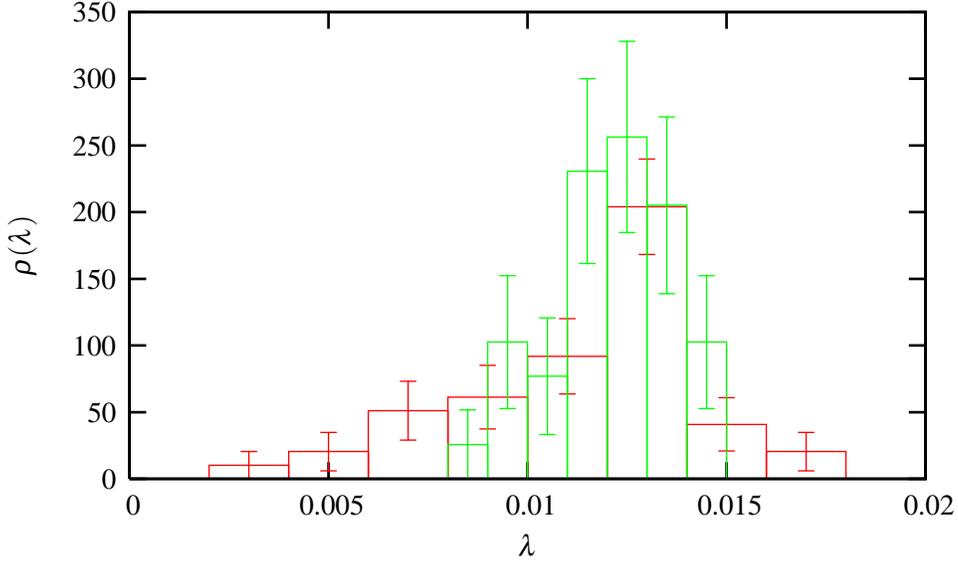}
\end{center}
\vspace{-5mm}
\caption{Eigenvalue density for the smallest eigenvalue of $\gamma_5D$ in $SU(3)$ at $\kappa=0.1557$~(red) and in $SU(6)$ at $\kappa=0.1581$~(green).}
\label{eig_fig}
\end{figure}

We have calculated the lowest Dirac eigenvalues at our lightest pion mass. We plot the distributions of eigenvalues we obtain in $SU(3)$ and $SU(6)$ in Fig.~\ref{eig_fig}. These are at similar pion masses, $am_\pi=0.188(6)$ and $0.212(3)$ respectively, but we see that the distribution of eigenvalues is much narrower in $SU(6)$, with no near--exceptional configurations. This suggests that it may be possible to use unimproved Wilson quarks at much lighter pion masses at large $N_c$ than is possible for $SU(3)$.

\begin{figure}[h]
\begin{center}
\input{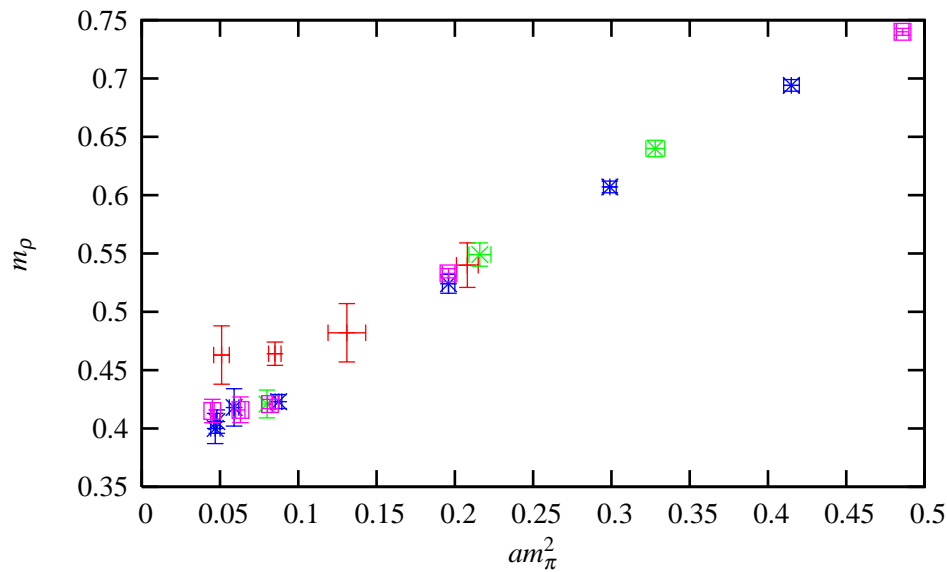}
\end{center}
\vspace{-5mm}
\caption{Rho mass against pion mass squared, for $SU(2)$~($+$), $SU(3)$~($\times$), $SU(4)$~($\ast$) and $SU(6)$~($\Box$).}
\label{mass_fig}
\end{figure}

We extract the ground state pseudoscalar and vector masses from correlators at a range of quark masses. 
The lowest pion mass is approximately $m_\pi=0.2a^{-1}=410$ MeV, corresponding to a quark mass somewhat lighter than the strange quark mass, while the heaviest is around $m_\pi=0.6a^{-1}=1220$ MeV. At our lighest masses $m_\pi/m_\rho\approx 0.5$. We show the extracted masses in Fig.~\ref{mass_fig}. We do not have exact matching of the pion masses for different $N_c$ in all cases, so we plot the rho masses against the squares of the pion masses --- if there is no $N_c$--dependence these should all fall on a common curve. We see that for the entire range of quark masses, the points do indeed fall onto the same line, the only exception being $SU(2)$, for which $m_\rho$ appears to be about 10\% higher at the lowest quark masses. Finite volume effects are expected to be largest at small $N_c$ and light quark masses, but we would expect these to increase $m_\pi^2$ by a larger amount than $m_\rho$, shifting the $SU(2)$ points below the line, the wrong direction to explain the observations. If the deviations are indeed real finite--$N_c$ effects, they are of order 10\%, similar to the deviations observed for glueballs in~\cite{LucTepWen}. Assuming they are dominated by the $1/N_c^2$ corrections, the corresponding deviations for $SU(3)$ should be around 4\%.

\section{Conclusions}
\label{conclusions}
We have calculated the masses of the pion and rho mesons up to $SU(6)$, at a cost approximately proportional to $N_c$. We have shown that the $N_c$ dependence of the masses is small, approximately 10\% for $SU(2)$ at small pion masses, and that there is no observable dependence at higher masses. If these differences are dominated by the $1/N^2$ corrections, which seems very plausible, there will be a difference of about 4\% between $SU(3)$ and $SU(\infty)$. Thus, at least for the masses of these mesons, quenched QCD is very close to the large--$N_c$ limit.

The distribution of the lowest eigenvalue of $\gamma_5D$ becomes much narrower as $N_c$ increases. This suggests that it maybe be possible to use Wilson quarks at significantly lower quark masses without running into exceptional configurations. We intend to explore this possibility in future work. We will also examine smaller volumes to check that the volume dependence does indeed decrease.

Apart from the ground state pseudoscalar and vector, we intend to also calculate the masses of the 
excited states. It would also be interesting to look at other quantum numbers, in particular the scalar mesons.

Finally we note that our calculations have been carried out at only one lattice spacing. Our results will thus be affected by lattice corrections. However, these lattice corrections will themselves have a large--$N_c$ limit, which will affect our results for all $N_c$ equally. Only an unlikely cancellation between the continuum and lattice $1/N_c^2$ corrections could cause us to see the very small $N_c$--dependence we observe. Thus, although our values for the meson masses have yet to be extrapolated to the continuum limit, our conclusion that the $N_c$--dependence is very small is unlikely to be affected. One should of course check this by repeating the calculations at a smaller lattice spacing, and we intend to do so in future.

\section*{Acknowledgments}
This work is supported by
the EC Hadron Physics I3 Contract RII3-CT-2004-506087
and by the GSI University Program Contract RSCHAE.

\end{document}